\documentclass[aps,prl,twocolumn,groupedaddress,floatfix,showpacs]{revtex4}

\usepackage{graphicx}
\usepackage{amsmath}

\begin{document}

\title{Quantum dots with Rashba spin-orbit coupling}

\author{M. Governale}

\affiliation{Institut f\"ur Theoretische Festk\"orperphysik,
Universit\"at Karlsruhe, D-76128 Karlsruhe, Germany}

\date{\today}

\begin{abstract}
We present results on the effects of spin-orbit coupling on 
the electronic structure of few-electron interacting quantum dots. 
The ground-state properties as a function of the number of 
electrons in the dot $N$ are calculated by means of spin density 
functional theory. We find a suppression of Hund's rule due 
to the competition of the Rashba effect and exchange interaction. 
Introducing an in-plane Zeeman field leads to a paramagnetic behavior of 
the dot in a closed shell configuration, and to spin texture in space.     
\end{abstract}


\maketitle
Spin-related phenomena have attracted great attention recently as  
 they are the key ingredient in the emerging field  of 
\emph{spintronics}\cite{wolf}. Among these, spin-orbit (SO) coupling 
mechanisms in semiconductors provide a basis for device applications, and 
a source of interesting physics, especially in systems with reduced 
dimensionality. 
Transport through chaotic quantum dots in the presence of SO interaction  
has been studied both experimentally\cite{folk} and 
theoretically\cite{halperin,alainer}, while the effect of SO coupling 
on the spin lifetime has been investigated in Ref.~\cite{nazarov}.  
Here, we are interested in how the electronic properties of 
few-electron quantum dots, such as the addition 
energy\cite{tarucha,macucci1} or the spin properties of the dot ground 
state\cite{koskinen},  are affected by Rashba SO 
interaction\cite{rashba, lommer}. 
These questions are interesting from the theoretical point of view for the 
following reasons.   
First, the Rashba effect has a different form than the usual 
SO coupling term in real atoms. Second, the 
tunability of the Rashba effect\cite{nitta,schaepers,grundler} allows 
dramatic SO effects to occur in quantum dots with few electrons;
in real atoms this requires heavy nuclei, and hence a more complicated 
electronic structure. 

We start by describing the physics of a quasi zero-dimensional system with 
Rashba SO at the non-interacting electron level, providing analytical 
results for the single-particle spectrum when the SO coupling can be 
treated as a perturbation. 
Then, we introduce the electron-electron interaction in the framework  
of Spin Density Functional Theory (SDFT)\cite{hedin}. This allows us to 
study how the addition spectrum is modified by varying  the strength of 
SO coupling. Studying the spin properties of 
the many-particle ground state, we find a suppression of Hund's rule, 
when the SO coupling can still be treated as a perturbation for 
the single-particle problem. For higher strengths it affects 
the single-particle spectrum so strongly  
that it gives rise to a completely different addition spectrum. 
The introduction of an in-plane magnetic field, 
leads to a paramagnetic behavior of the dot in a closed shell configuration, and to spin texture in space.   

Quantum dots are often  realized by lateral confinement of a 
two dimensional electron gas (2DEG) obtained in a heterostructure. 
Due to the lack of inversion symmetry along the growth 
direction $z$ of the heterostructure\cite{rashba, lommer}, 
the electrons in the 2DEG are subject to the 
Rashba spin-orbit coupling Hamiltonian 
\begin{equation}
\label{hso}
H_{\text{so}} = \frac{\hbar k_{\text{so}}}{m}\left( \sigma_x\,
p_y - \sigma_y\, p_x\right)\quad.  
\end{equation}
The strength of the SO coupling, here denoted as $k_{\text{so}}$,  
can be tuned by changing the asymmetry of the quantum well via 
externally applied voltages, as shown in several experimental 
studies\cite{nitta,schaepers,grundler}. 

It is interesting to study the effect of the SO  
coupling term Eq.~(\ref{hso}) on the quantum mechanics of a 
quasi zero-dimensional system\cite{voskob}. To this end, 
we consider a two-dimensional 
quantum dot defined by a parabolic confining potential
\begin{equation}
\label{conf}
V_{\text{conf}}(x,y) = \frac{m}{2}\,\omega^2\, (x^2+y^2)\quad. 
\end{equation}
Thus,  the single-particle Hamiltonian in the effective-mass approximation 
reads
\begin{equation}
H=\frac{p_x^2+p_y^2}{2 m}+ V_{\text{conf}}(x,y)+H_{\text{so}}\quad.
\end{equation}
In the absence of SO coupling the eigen-energies are
\begin{equation}
\label{e0}
E_{M}^{(0)}=\hbar \omega \left(M+1\right),   
\end{equation}
with $M$ being a non-negative integer. 
A degenerate subspace $\mathcal{S}_M$ of dimension $D_M=2(M+1)$, 
where the factor $2$ is due to spin, is associated to each energy $E_{M}^{(0)}$. 

\begin{figure}
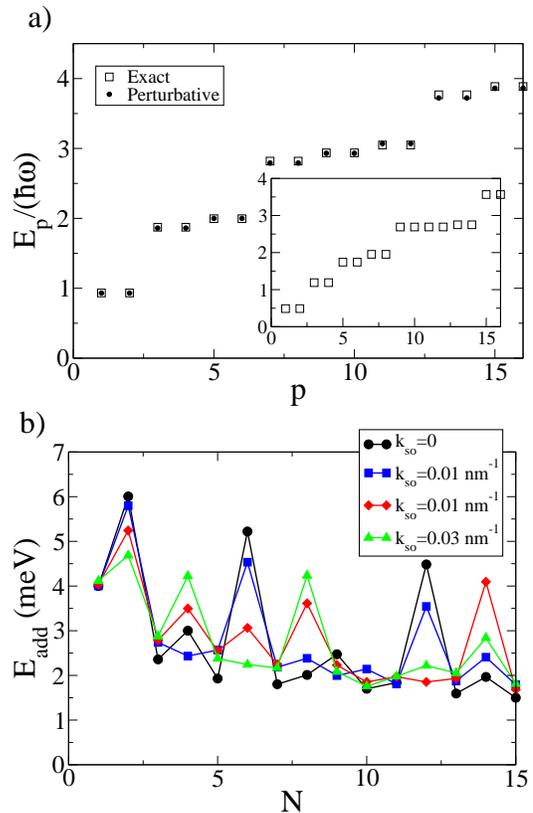

\includegraphics[width=2.7in]{fig1a.eps}\\
\includegraphics[width=2.7in]{fig1b.eps}
\caption{
(a) Low-energy part of the single-particle spectrum for the dot, calculated 
both by perturbation theory (dots) and by numerical diagonalization 
(squares) for $k_{\text{so}} l_\omega=0.2633$ [this value corresponds to 
the data for $k_{\text{so}}=0.01\ {\text{nm}}^{-1}$ for the addition energies 
of the realistic dot shown in panel (b)].  
The label $p$ is an index that enumerates the eigenstates 
in order of ascending energy.
Inset: Low-energy part of the 
single-particle  
spectrum calculated numerically for $k_{\text{so}} l_\omega=0.7896$ 
[this value correspond to $k_{\text{so}}=0.03\ {\text{nm}}^{-1}$ in panel (b)]
In this case the SO coupling dominates the single-particle spectrum.
\\
(b)  Addition energy vs number of electrons in the dot for 
different values of the SO coupling strength. In this figure 
and in the following ones the dot is 
defined by a confining potential of strength $\hbar \omega=5 {\text{ meV}}$. 
\label{addition}}
\end{figure}

We will now treat $H_{\text{so}}$ as a perturbation. 
This is valid as long as $ k_{\text{so}}l_{\omega}\ll 1$, where 
$l_\omega$ is the oscillator length $\sqrt{\hbar/(m \omega)}$. 
We obtain for the second-order eigen-energies  
\begin{equation}
\label{epert}
\tilde{E}_{M,i,\sigma}=E_M^{(0)}+\hbar \omega (k_{\text{so}}
l_\omega)^2\left[2 (i-1)-(M+1)\right],  
\end{equation}
where $i=1,\cdots,M+1$, and $\sigma=\pm 1$ is the quantum number relative 
to $\sigma_z$, i.e. the spin projection along $z$. 
As the single-particle levels will play an important role in the 
following, we show an example of the low-energy part of the 
spectrum calculated using Eq.~(\ref{epert}) 
together with the results of numerical diagonalization
in the upper panel of Fig.~\ref{addition}. 
From the perturbative treatment the following conclusions can be 
drawn: 
1) Each degenerate level  $E_M^{(0)}$ is split in $M$ sublevels, each 
of which is double degenerate due to Kramers theorem;
2) Spin rotational invariance is broken but still the eigenstates 
are (to this order in perturbation theory) eigenstates of $\sigma_z$. 
From conclusion 1) we can infer that SO coupling changes the 
addition spectrum of the dot, while conclusion 2) tells us that 
the Rashba effect will not influence the lifetime of the eigen-states 
of $\sigma_z$.  
For values of $k_{\text{so}}$ for which  
perturbation theory breaks down, the eigen-energies 
are of course still grouped in Kramers-degenerate 
sublevels [as shown in the inset of Fig.~\ref{addition}(a)], 
although it can happen that different sublevels have almost the same energy.

To introduce the Coulomb interaction between the electrons  we 
use spin density functional theory (SDFT), in the local density 
approximation\cite{hedin}. 
We write the Kohn-Sham equation\cite{kohn1} in spinor notation
\begin{eqnarray}
\label{kohn}
\nonumber
& &\left[\left(-\frac{\hbar^2}{2m}\nabla^2+V_{\text{conf}}+
V_{\text{coul}}\right)\mathbf{1}+
H_{\text{so}}+
\mathbf{V_{\text{exch-corr}}}\right] \Phi_j
\\& &=\varepsilon_j \Phi_j,  
\end{eqnarray}
where $V_{\text{coul}}$ is the Hartree potential, 
$\mathbf{V_{\text{exch-corr}}}$ the exchange-correlation potential, 
$\mathbf{1}$ the identity matrix in spin space, and  
$\Phi_j$ is a two-components spinor. 
The spin-dependent exchange-correlation potential is introduced following 
Ref.~\cite{hedin}. It is in general non diagonal in spin space 
(it becomes diagonal in the basis that diagonalizes the spin-density 
matrix).  For the exchange-correlation energy we use  the 
Tanatar and Ceperly parameterized form\cite{tanatar}; in 
particular for the case of partial spin polarization we use  
the interpolation scheme of Refs.~\cite{hedin,koskinen}. 
We solve Eq.~(\ref{kohn})  self-consistently by discretizing it in real space. 
The addition energy 
(also called capacitive energy by some 
authors\cite{macucci1}) is 
defined as $E_{\text{add}}=\mu(N+1)-\mu(N)$, where $\mu(N)$ is the 
chemical potential for the dot with $N$ electrons, i.e. the energy needed 
to add the $N$-th electron to the system containing already $N-1$ electrons.
We compute the chemical potential by means of Slater's 
rule\cite{perdew1}, in order to minimize numerical errors due to double 
differentiation.

Now, we focus on  realistic dots obtained in an InAs heterostructure, where 
the Rashba effect can be quite large\cite{grundler}.  
We use for the electron effective mass the value $m=0.022 \  m_0$, with $m_0$ 
being the free-electron mass; and for the dielectric constant $\epsilon= 14.6
\ \epsilon_0$, being $\epsilon_0$ the one of vacuum.
   
In a quantum dot without SO interaction we expect peaks in 
the addition energy when the number of electrons $N$ equals a 
\emph{magic numbers}, i.e at those integer 
values which correspond to a closed shell configuration. 
For a parabolic dot 
the first magic numbers are 2,6,12, see Eq.~(\ref{e0}) and the 
discussion below it. 
Besides these peaks for $N$ coinciding 
with a magic number, some additional peaks are expected for a number of 
electrons corresponding to a half-filled shell due 
to Hund's rule. In this situation the electrons in the half-filled 
shell have parallel spins to gain exchange energy.   
For the parabolic dot under consideration the first Hund's-rule peaks are 
located at $N=4$, and $N=9$. Both these kind of peaks can be seen in 
Fig.~\ref{addition}(b), for the case when no SO coupling is present
(filled dots). 
Switching on the SO interaction leads to a change in the single-electron 
levels of the dot, and such a change is reflected in the addition 
energy. This effect can be seen in Fig.~\ref{addition}(b), where the 
addition energy is plotted for several values of the SO coupling 
strength, ranging from a situation where the perturbative treatment 
Eq.~(\ref{epert}) is still valid, to one where the SO coupling dominates 
the single-particle spectrum. 
Due to the presence of Kramers-degenerate sublevels in the free-electron 
spectrum, peaks tend to be present for even number of 
electrons in the dot.    
The fact that a degenerate level $E_M^{(0)}$ 
is split in $M$ Kramers-degenerate sublevels (in the perturbative regime), 
leads to a suppression of the Hund's rule: in a half-filled 
level to maximize the total spin the electrons should be allocated 
one per sublevel, but this has an energy cost equal to the  sublevel  
splitting; if the sublevel splitting is larger than the gain in exchange 
energy, then the spin polarization for a half-filled level is 
suppressed. This is indeed what we see by analyzing the 
spin properties of the SDFT ground-state wave-function. 
At this point it is important to stress that due to the 
tunability of the SO coupling strength\cite{nitta, schaepers, grundler}, 
it is possible to investigate experimentally the effect of the Rashba 
term on the addition spectrum of few-electrons quantum dots
(addition spectra were measured by Tarucha \emph{et al.}\cite{tarucha}), 
and the transition from weak ($k_{\text{so}}l_\omega \ll1$) to strong 
SO coupling. 
  
We now investigate the effect of an in-plane magnetic field. 
Due to the fact that the system is invariant under rotation around the 
$z$-axis, we can choose the direction of the in plane magnetic field 
arbitrarily without losing any generality.  
We introduce a magnetic field $B$ along $x$, which 
does not affect directly the orbital motion, but couples to the 
$x$-component of the total spin,  
giving rise to a Zeeman term, $H_z=\hbar \omega_z S_x/2$, 
where $S_x=\sum_{i=1,N} \sigma_x^{(i)}$, and $\hbar \omega_z=\mu_{\text{B}} 
g^* B$, 
with $\mu_{\text{B}}$ being Bohr's magneton, and $g^*$ the  g-factor. 
\begin{figure}
\includegraphics[width=2.8in]{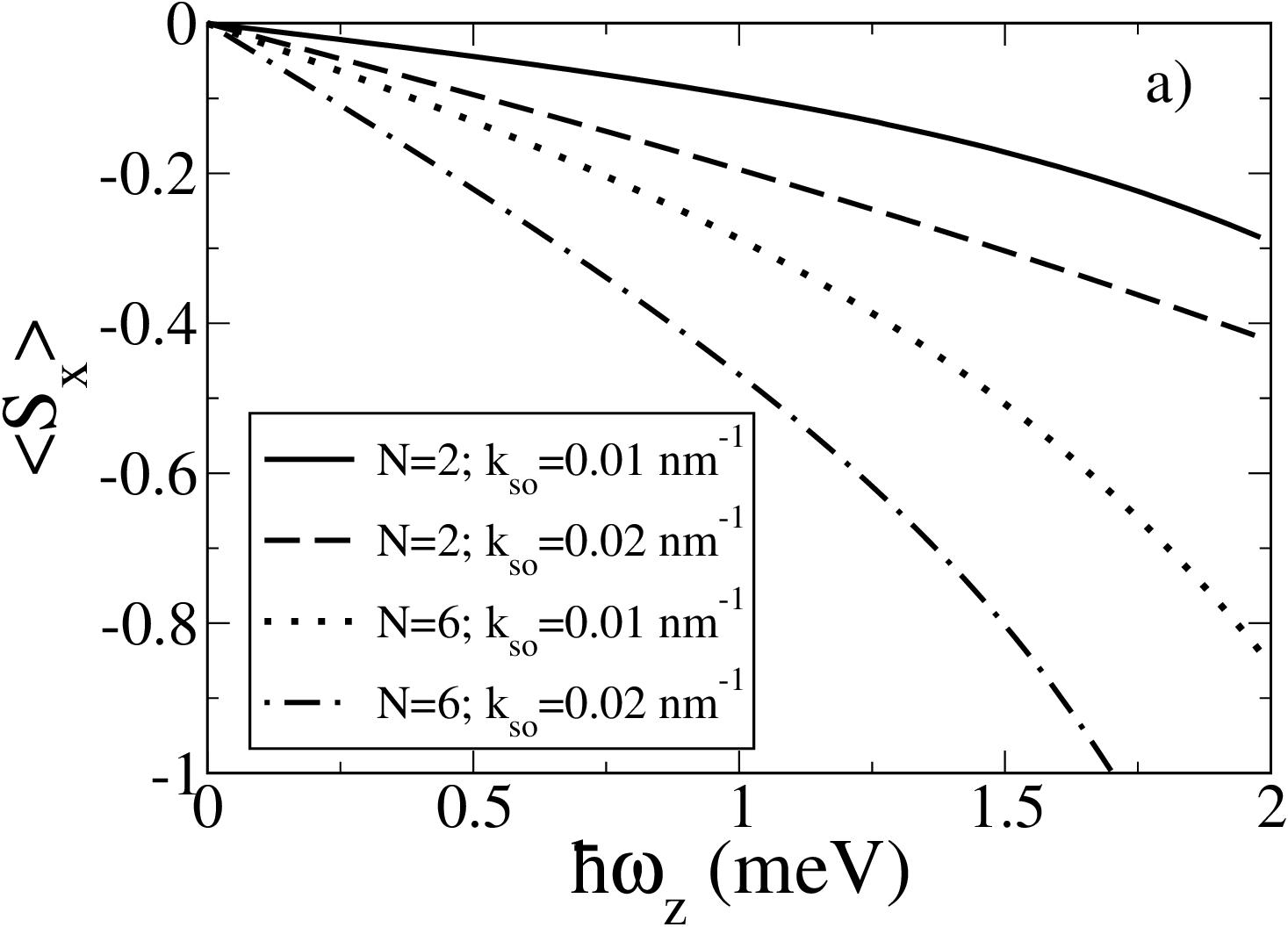}\\
\includegraphics[width=2.7in]{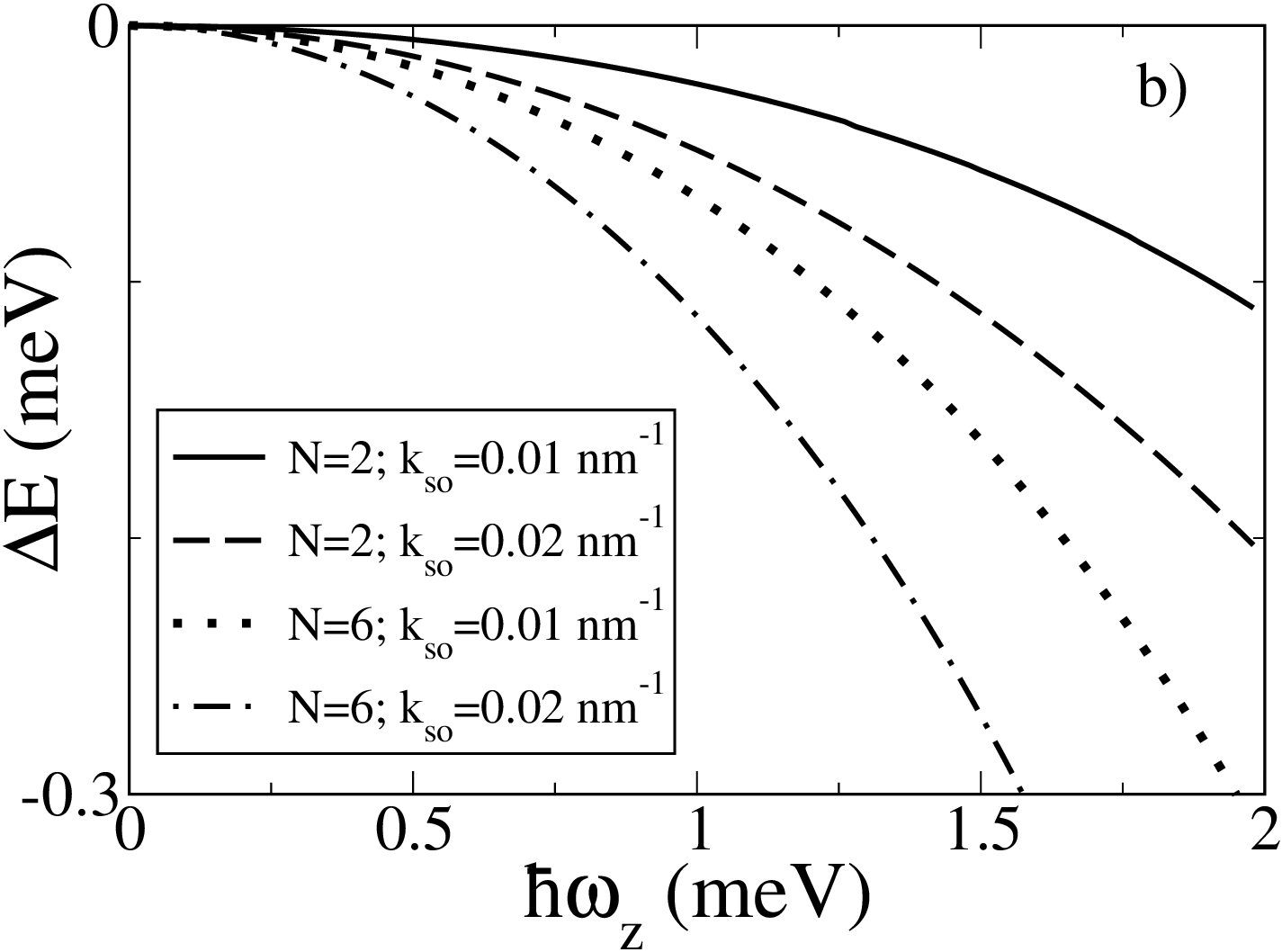}
\caption{ 
(a) Average values of $S_x$ vs the strength of the in-plane magnetic field, 
computed by means of SDFT,  
for two closed shell configurations 
($N=2$ and $N=6$) and for different strength of the spin-orbit interaction. 
Without the Rashba term, $\langle S_x\rangle$ would be zero.\\  
(b) Variation of the ground state energy $\Delta E= E(B)-E(0)$  
vs the strength of the in-plane magnetic field, computed by means of SDFT. 
The ground state energy shows a quadratic dependence on the in-plane magnetic 
field:
$E(B)=E(0)-A B^2$, with $A>0$. 
The susceptibility is positive and the 
system is paramagnetic. 
\label{magnetic}}
\end{figure}
We consider now a dot in a closed-shell configuration, namely we take 
$N=2$ and $N=6$. In the case of vanishing SO coupling and in 
the independent-electron picture, such a 
system does not respond to the in-plane magnetic field 
for Zeeman splitting smaller 
than the level splitting ($\omega_z <\omega$).  
The situation changes when the Rashba term is introduced; the ground state 
of the dot 
exhibits now some net-spin polarization. 
In the upper panel of Fig.~\ref{magnetic}, 
the average value of $S_x$ is plotted vs magnetic field, showing 
how the system gets magnetized even in a closed-shell configuration due 
to the interplay of SO coupling and Zeeman splitting. 
The average values of $S_y$ and $S_z$ remain equal to zero.  
In the lower panel of Fig.~\ref{magnetic}, 
the variation of the ground state energy with magnetic field 
is plotted vs magnetic field. It shows a decrease with 
increasing field (which is  well fitted by a parabola), 
yielding a positive susceptibility $\chi=-\partial^2 E 
/\partial B^2$. Thus, we can conclude that the dot in a closed shell 
configuration exhibits a \emph{paramagnetic} behavior. 
This is in contrast to what happens in real atoms, where a 
closed shell gives a diamagnetic response due to orbital 
degrees of freedom (Larmor diamagnetism)\cite{vanvleck}, while in 
our case the Larmor term 
is suppressed by the two-dimensionality of the dot.
This paramagnetic behavior\cite{rashbapar} is due to the single-particle 
eigen-states (see below), 
but it persists when the electron-electron interaction is 
present (it is enhanced by it), as shown in Fig.~\ref{magnetic}.

In the limit of $k_{\text{so}} l_\omega \omega /\omega_z \ll 1$, 
and $w_z<w$, it is possible to 
obtain a perturbative expression in $H_{\text{so}}$ 
for the single-particle eigen-energies:
\begin{eqnarray}
\nonumber
\tilde{E}_{M,i,\sigma}&=&E_M^{(0)}+\sigma\frac{\hbar\omega_z}{2}
-\frac{\hbar \omega}{2} (k_{\text{so}} l_\omega)^2 \\
& &\left\{1+\frac{\omega^2}{\omega^2- \omega_z^2}\left[1+\sigma 
(2 i-1) \frac{\omega_z}{\omega}\right]\right\},
\end{eqnarray}
where $i=1\cdots M+1$, $\sigma=\pm 1$ is the quantum number relative to 
the projection of spin 
in the direction of the magnetic field, i.e $\sigma_x$, and $E_M^{(0)}$ are the 
energies given in Eq.~(\ref{e0}). 
In the independent-electron approximation, the energy of a closed 
shell $E_{\text{cs}}(M)=\sum_{i,\sigma}\tilde{E}_
{M,i,\sigma}$ reads 
\begin{eqnarray}
\nonumber  
 E_{\text{cs}}(M)&=& 2(M+1)E_M^{(0)}\\
& & -(M+1)\hbar\omega(k_{\text{so}}l_\omega)^2 \left[1+\frac{\omega^2}
{\omega^2-\omega_z^2}\right].
\label{ecs}
\end{eqnarray}
Expanding Eq.~(\ref{ecs}) in $\omega_z/\omega$ we get for the 
magnetic-field dependent part of the closed-shell energy 
$-(M+1)\hbar \omega (k_{\text{so}}l_\omega)^2 (\omega_z/\omega)^2+
\mathcal{O}[(\omega_z/\omega)^4]$, which explains the parabolic behavior 
seen in Fig.~\ref{magnetic}. 
From Eq.~(\ref{ecs}) we get a paramagnetic 
contribution to the susceptibility due to a closed-shell 
\begin{equation}
\frac{\chi_M}{(g^* \mu_{\text{B}})^2} =2 (M+1)\omega^3 
(k_{\text{so}}l_\omega)^2 \frac{\omega^4-\omega_z^4}{\hbar
\left(\omega^2-\omega_z^2\right)^4},
\end{equation}
which in the limit of $\omega_z/\omega\ll 1$ is just a positive constant: 
$\chi_M/(g^* \mu_{\text{B}})^2= 
2 (M+1)(k_{\text{so}}l_\omega)^2/ (\hbar \omega)$.

\begin{figure}
\includegraphics[width=2.7in]{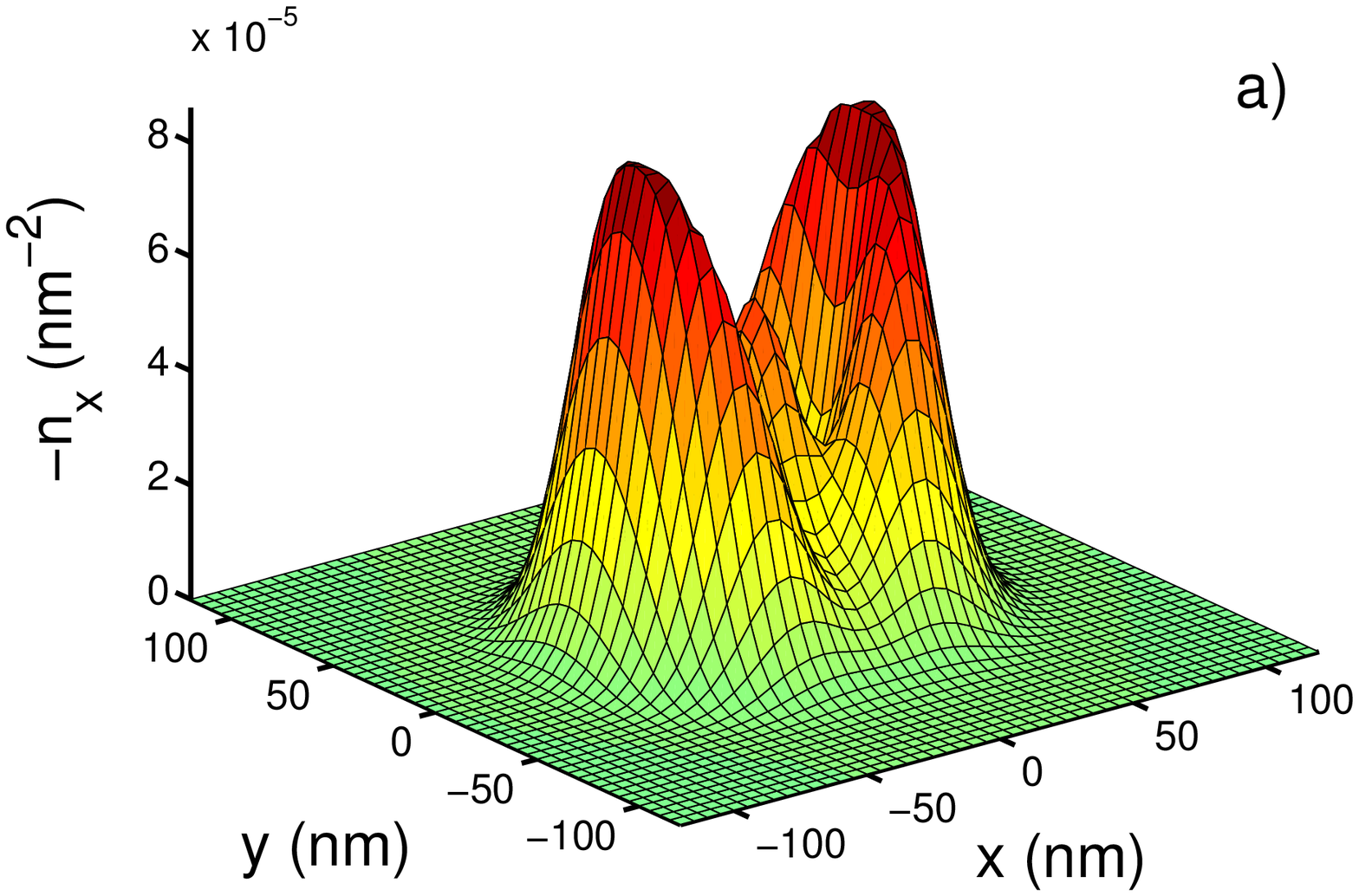}\\
\includegraphics[width=2.7in]{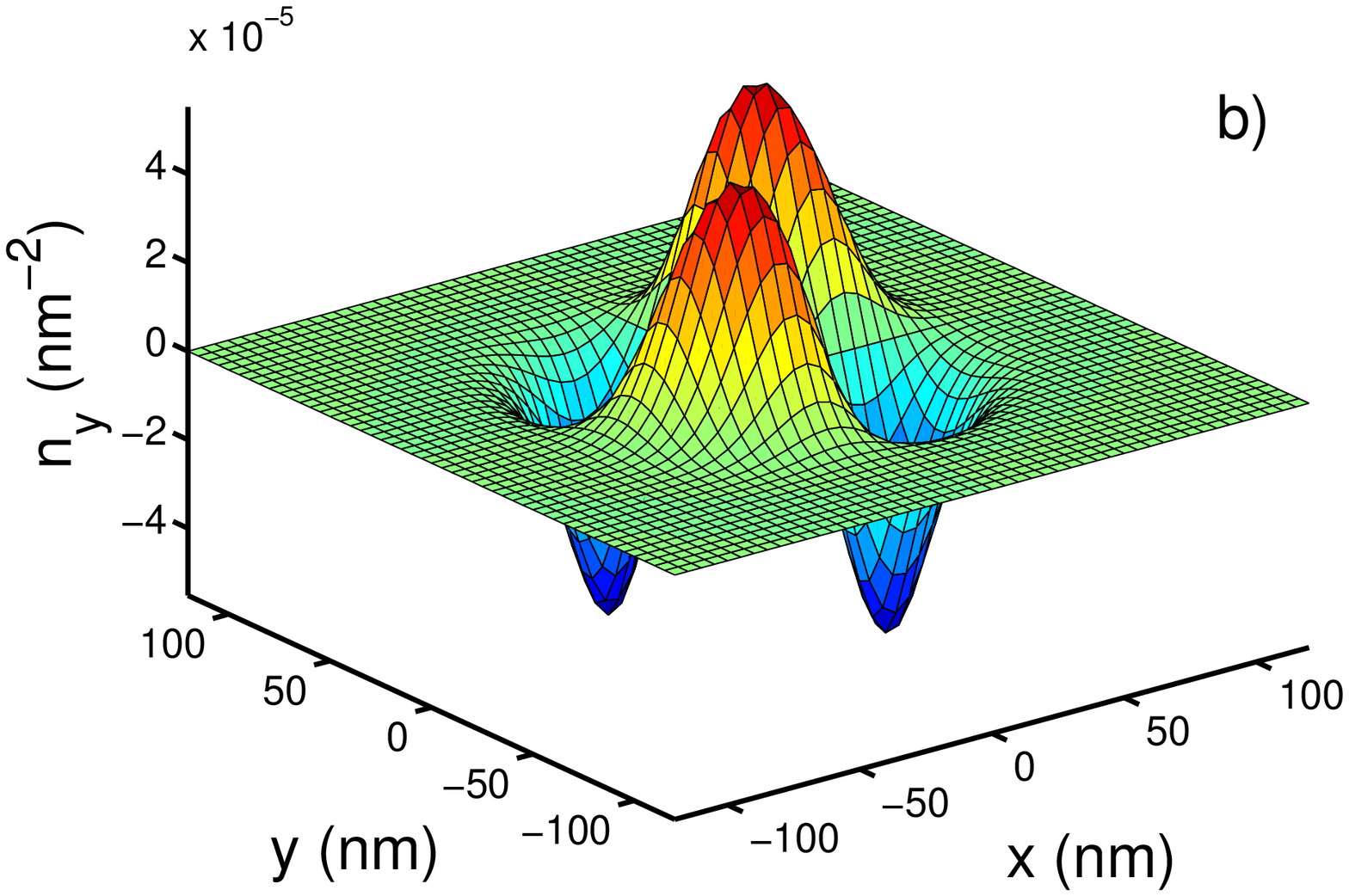}\\
\includegraphics[width=2.7in]{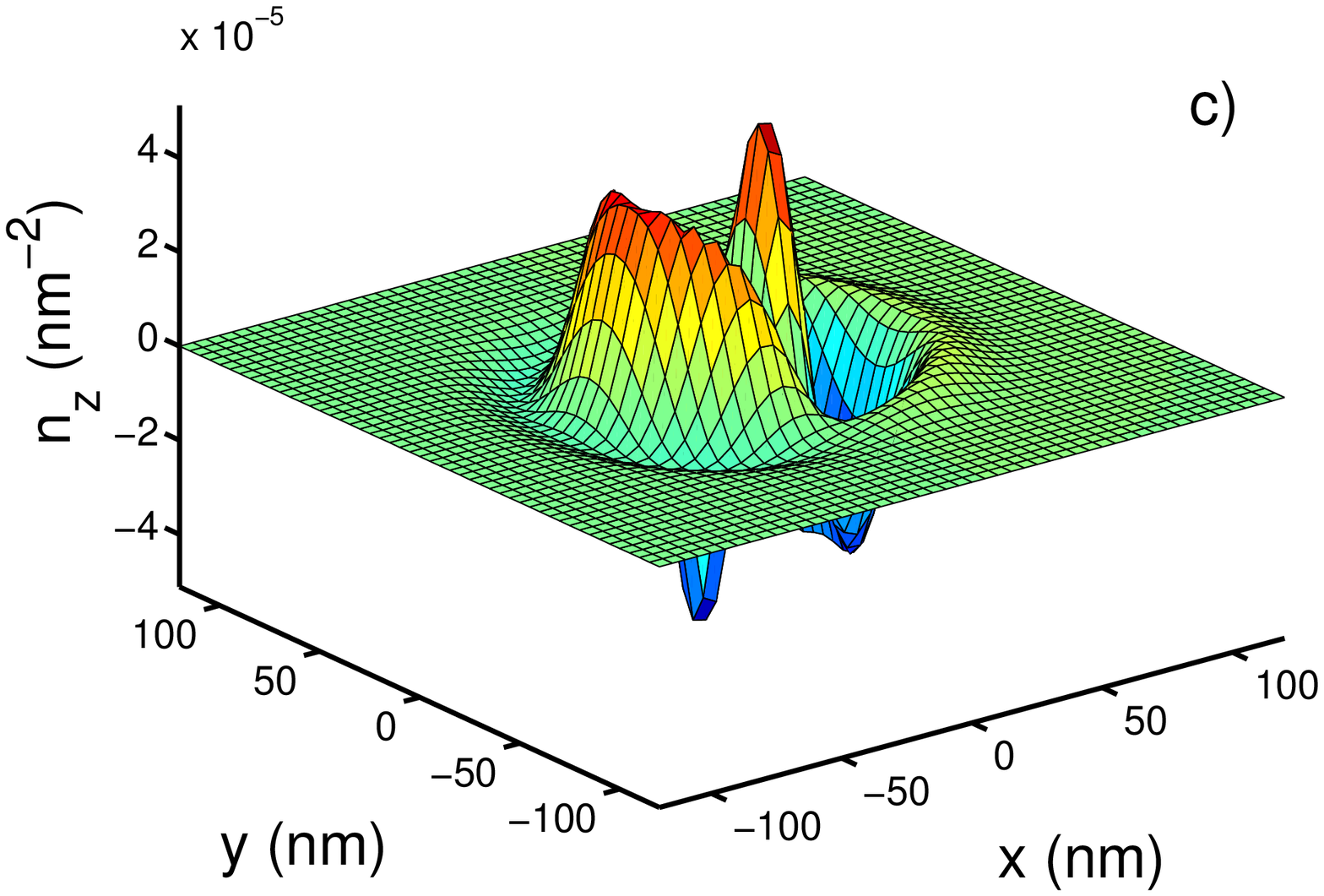}
\caption{Spin density projected along $x$ (a), $y$ (b), and $z$ (c), 
computed by means of SDFT,  
for a dot with $N=6$ electrons, $k_{\text{so}}=0.02 {\text{ nm}}^{-1}$, and 
$\hbar \omega_z=1 {\text{ meV}}$. In (a), for ease of visualization, we plot 
$-n_x$ instead of $n_x$. Spatial integration of $n_y$ and $n_z$ gives 
zero, yielding a zero average for the corresponding total-spin components.   
\label{dens}}
\end{figure}
      
It is interesting to have a closer look at the spin-density  for 
the magnetized dot. In Fig.~\ref{dens} 
the projections of the spin density along the $x$-, $y$-, and $z$-axis are 
shown for a dot containing six electrons in the presence both of 
SO coupling and of a Zeeman field. As it is clearly visible,  
the system shows spin texture in space, this is due to the fact that 
no common spin-quantization axis exists anymore
(a similar situation occurs 
in quantum wires with strong spin-orbit coupling\cite{rashbawire}.) 

In conclusion, we have studied the effect of Rashba spin-orbit interaction 
on the addition energy, and on the spin properties of a few-electron 
quantum dot by means of spin density functional theory. 
In particular, we have found a suppression of Hund's rule, for small
$k_{\text{so}}$ values, for which perturbation theory in $H_{\text{so}}$ still 
holds. 
An additional in-plane magnetic field 
(Zeeman field) leads to a paramagnetic behavior of the dot in 
a closed shell configuration, and to spin texture in space.       

This work was supported by the Center for Functional 
Nanostructures at the University of Karlsruhe. We thank J.~C. Cuevas, 
M.~Macucci, and  U.~Z\"ulicke for useful discussions.


\begin{thebibliography}{24}
\bibitem{vanvleck} For a closed atomic shell 
Van Vleck paramagnetism is absent.
\end{thebibliography}
\end{document}